\documentclass[aps,pra, preprint,showpacs]{revtex4-1}  

\usepackage[citecolor=blue,colorlinks=true]{hyperref}
\usepackage{hyperref}
\usepackage[english]{babel} 
\usepackage{graphicx} 
\usepackage{epstopdf} 
\usepackage{amssymb, amsmath} 

\usepackage{textpos}
\begin{document}
\begin{textblock*}{200mm}(.01\textwidth,-2.2cm)
\large \textcolor{blue}{TO APPEAR IN PHYSICAL REVIEW A}
\end{textblock*}

\title{An N-atom Collective State Atomic Clock with Root-N Fold Increase in Effective Frequency and Root-N Fold Reduction in Fringe Width}

\author{May E. Kim}
\email{mekim@u.northwestern.edu}
\affiliation{Department of Physics and Astronomy, Northwestern University, Evanston, IL, USA 60208}

\author{Resham Sarkar}
\affiliation{Department of Physics and Astronomy, Northwestern University, Evanston, IL, USA 60208}

\author{Renpeng Fang}
\affiliation{Department of Physics and Astronomy, Northwestern University, Evanston, IL, USA 60208}

\author{Selim M. Shahriar}
\affiliation{Department of Electrical Engineering and Computer Science, Northwestern University, Evanston, IL, USA 60208}
\affiliation{Department of Physics and Astronomy, Northwestern University, Evanston, IL, USA 60208}

\date{\today}

\begin{abstract}
We describe a collective state atomic clock (COSAC) with Ramsey fringes narrowed by a factor of $\sqrt{N}$ compared to a conventional clock -- $N$ being the number of non-interacting atoms -- without violating the uncertainty relation. This narrowing is explained as being due to interferences among the collective states, representing an effective $\sqrt{N}$ fold increase in the clock frequency, without entanglement. We discuss the experimental inhomogeneities that affect the signal and show that experimental parameters can be adjusted to produce a near ideal signal. The detection process collects fluorescence through stimulated Raman scattering of Stokes photons, which emits photons predominantly in the direction of the probe beam for a high enough optical density. By using a null measurement scheme, in which detection of zero photons corresponds to the system being in a single collective state, we detect the population in a collective state of interest. The quantum and classical noise of the ideal COSAC is still limited by the standard quantum limit and performs only as well as the conventional clock. However, when detection efficiency and collection efficiency are taken into account, the detection scheme of the COSAC increases the quantum efficiency of detection significantly in comparison to a typical conventional clock employing fluorescence detection, yielding a net improvement in stability by as much as a factor of 10. 
\end{abstract}

%
\maketitle

\section{Introduction}		
It is well known that the width of the fringes, observed as a function of the detuning, in a pulsed excitation of an atomic transition, is limited by the inverse of the interaction time. This effect is routinely observed in systems such as microwave or Raman atomic clocks \cite{niering,santarelli,arimondo,esnault,wilpers}. It is also well known that the effective interaction time can be extended by employing Ramsey's technique of separated field excitations \cite{ramsey}. In that case, the transit time limited linewidth is determined by the inverse of the time delay between the two fields. The temporal profile of the field envelope seen by the atoms is a pair of square pulses, each with a duration $T_1$, separated by $T_2$. For a conventional clock (CC), the Ramsey technique produces a sync function with a width of $\sim T_1^{-1}$, modulated by a sinusoid with a fringe width of $\sim T_2^{-1}$, all centered at the carrier frequency.

The width of these fringes can be reduced by making use of entanglement, as demonstrated by Wineland et al. using trapped ions \cite{wineland00}.  Consider, for example, a situation where the use of entanglement allows one to couple the ground state of three particles to a state where all three  particles are in the excited state, representing a collective excitation.  This corresponds to an effective increase in the transition frequency by a factor of three.  As such, the detuning for a single atom gets tripled for this collective excitation, so that the width of the Ramsey fringe gets reduced by a factor of three.  However, realizing such a scheme for a large number of particles is beyond the capability of current technology.

Here, we describe a scheme that produces Ramsey fringes that are narrower by a factor of more than $10^3$ for parameters that are readily accessible, without making use of entanglement. While the concept can be applied to other types of atomic clocks, as described later, the specific experiment we propose is an optically off-resonant Raman atomic clock using ensembles of $N$ cold atoms. The clock transition is detected by measuring one of the collective states rather than measuring individual atomic states. The fringes observed as a function of the Raman (i.e. two photon) detuning is found to be $\sim \sqrt{N}$ times narrower than the transit time limited width that would be seen by measuring individual atomic states, as is the case with the CC. For the current state of the art of trapped atoms, the value of $N$ can easily exceed $10^6$, so that a reduction of fringe width by a factor of more than $10^3$ is feasible.   

The reduction in the width of the fringe, especially by such a large factor, strongly violates the conventional transit time limit of spectroscopic resolution.  However, we show, via a detailed analysis of the standard quantum limit and the Heisenberg limit, that, indeed, this violation of the conventional transit-time limit is allowed, and is within the constraint of the more fundamental uncertainty principle of quantum mechanics.  We also show that under certain conditions, frequency fluctuation of the COSAC can be significantly smaller, by as much as a factor of 10, than that for a fluorescence detection based conventional clock employing the same transition and same atomic flux. The ultra-narrow resonances produced in this process may also open up the possibility of exploring novel ways of implementing spin-squeezing techniques for further improvement in clock stability  \cite{wineland01,kitagawa,hald,kuzmich}. 

The rest of the paper is organized as follows. In Section \ref{sec:three}, we introduce a single three level atomic system and how it propagates through a Ramsey fringe experiment. In Section \ref{sec:collective}, we derive the propagation of a collective state through the same Ramsey fringe experiment, showing mathematically the narrowing of the fringe by $\sqrt{N}$. In the subsections, the effects of velocity distribution, field inhomogeneity, spontaneous emission, and fluctuation in the number of atoms are discussed. We show that while these effects tend to degrade the signal, these limitations can be circumvented with proper choice of experimental parameters. In Section \ref{sec:experiment}, we lay out the scheme for realizing the COSAC experimentally. The detection scheme is fundamentally different from that of the CC since only a single collective state is detected. Because the atoms are in a superposition of collective states at the end of the Ramsey fringe experiment, and the CC detects signal from one level of the (reduced) two level system, such detection scheme collects signal from most of the collective states. In contrast, the heterodyne detection scheme employed for the COSAC ensures that only a single collective state is detected. In Section \ref{sec:performance}, the performance of the COSAC is compared to that of the CC by analyzing quantum and classical noise, detector efficiency, and collection efficiency. In Section \ref{sec:interpret}, we present the physical interpretation for why the linewidth narrows for a COSAC. We ensure, by proper interpretation of the frequency uncertainty and observation time, that the fundamental quantum limit is not violated. Lastly, in Section \ref{sec:conclusion}, we conclude with a summary of the paper.

		
\section{Three level atomic system in Ramsey fringe experiment}
\label{sec:three}

The optically off-resonant Raman atomic clock employs three hyperfine energy levels in a $\Lambda$ scheme depicted in Fig. \ref{fig:pop} (a). The ground states $|1\rangle$ and $|2\rangle$ of this atom interact with an excited state $|3\rangle$ via two coherent electromagnetic light fields of frequencies $\omega_1$ and $\omega_2$, respectively, detuned from resonance by $\delta_1$ and $\delta_2$, respectively. The Hamiltonian after the dipole approximation, rotating wave approximation, and rotating wave transformation can be expressed as \cite{shahriar00}:
\begin{equation}
\displaystyle
H=\frac{\hbar}{2}\left[(\delta\sigma_{11}-\delta\sigma_{22} -2\Delta\sigma_{33})-\left(\Omega_1\sigma_{13}+\Omega_2\sigma_{23}+h.c.\right)\right]
\end{equation}
where $\sigma_{\mu\nu}=|\mu\rangle\langle \nu|$, $\delta \equiv \delta_1-\delta_2$ is the two photon detuning, $\Delta \equiv (\delta_1+\delta_2)/2$ is the average detuning, and $\Omega_{1,2}$ are the Rabi frequencies. Here, we have also assumed a phase transformation applied to the Hamiltonian so that $\Omega_{1,2}$ are real. We assume next that $\Delta\gg\Gamma,\Omega_1$, and $\Omega_2$ (where $\Gamma$ is the decay rate of state $|3\rangle$) so that the effect of $\Gamma$ can be neglected, and state $|3\rangle$ can be eliminated adiabatically \cite{shahriar01,kasevich} (in Section \ref{sec:collective}, we will consider the residual effect of spontaneous emission). Under these conditions, the Hamiltonian of the reduced two level system can be expressed as $H_{red}=(\hbar\delta/2)\sigma_z-(\hbar\Omega/2)\sigma_x$, where $\Omega \equiv \Omega_1 \Omega_2/2\Delta$ is the Raman Rabi frequency, and $\sigma_z$ and $\sigma_x$ are Pauli matrices defined as $\sigma_z=(\sigma_{11}-\sigma_{22})$ and $\sigma_x=(\sigma_{12}+\sigma_{21})$. The quantum state for this system is given by $|\psi(t'+t)\rangle = W^{\delta t}_{\Omega t}|\psi (t')\rangle$ where $|\psi(t')\rangle=\tilde{c}_1(t')|1\rangle+\tilde{c}_2(t')|2\rangle$, and the propagation operator is given by \cite{scully}
\begin{equation}
W^{\delta t}_{\Omega t}=e^{i\delta t/2}
	\begin{pmatrix}
		\cos\phi-i \frac{\delta}{\Omega'}  \sin\phi &-i\frac{\Omega}{\Omega'}\sin\phi \\
		-i\frac{\Omega}{\Omega'}\sin\phi
		& \cos\phi+i \frac{\delta}{\Omega'}  \sin\phi
	\end{pmatrix}
	\label{propagation}
\end{equation}
where $\phi=\Omega' t/2$, and $\Omega' \equiv \sqrt{\Omega^2+\delta^2}$ is the generalized Rabi frequency. 


When this system is excited by two pulses of duration $T_1$, 
separated in time by $T_2$, we have $\Omega_1(t)\sim\Omega_2(t)=\Omega_0 [ U(t)-U(t-T_1)+U(t-(T_1+T_2))-U(t-(2T_1+T_2)) ]$ where $U(t)$ is the Heaviside step function. When $\delta\ll\Omega$ and the width of the pulse is chosen to be $\Omega T_1=\pi/2$, each pulse acts on the system as a propagation operator $W^0_{\pi/2}=(I-i\sigma_x)/\sqrt{2}$. While the system is between $t=T_1$ and $t=T_1+T_2$ where no interaction is present, the propagation operator can be expressed as $W^{\delta T_2}_0=\sigma_{11}+e^{i\delta T_2}\sigma_{22}$. After passing through the three zones, the state of the atom that was originally in state $|1\rangle$ is $|\psi\rangle=W^0_{\pi/2}W^{\delta T_2}_0W^0_{\pi/2}|1\rangle=-i e^{i\theta}(\sin{\theta}|1\rangle +\cos{\theta}|2\rangle)$ where $\theta=\delta T_2/2$ is the dephasing angle. The probability of the atom being in state $|2\rangle$ is $P_2\equiv|\langle 2|\psi\rangle|^2=(\cos{\theta})^2$.


\section{Collective state atomic system in Ramsey fringe experiment}
\label{sec:collective}

The discussion can be generalized to $N$ atoms that are all excited by the same field. We assume that there are no overlaps between the wavefunctions of the atoms and there is no interaction among them \cite{dicke}. The evolution of each atom under these assumptions can be described individually, and the total quantum state is simply the outer (tensor) product of individual quantum states \cite{shahriar02,arecchi}. However, the interaction can also be described equivalently using a basis of collective states \cite{shahriar02,dicke}. The Hilbert space of N two level atoms is spanned by $2^N$ states. Thus, when transformed to the collective state basis, there are also $2^N$ collective states. For identical Rabi frequencies and resonant frequencies, however, only the generalized symmetric states \cite{shahriar02}, of which there are only $(N+1)$, are relevant, and the rest of the $(2^N-N-1)$ states become decoupled. The case where inhomogeneity of the Rabi frequencies and different Doppler shifts experienced by different atoms are taken into account is presented at the end of this section. We also note that if different atoms see different phase factors from the excitation fields, these factors can be absorbed into the definition of the generalized symmetric states \cite{shahriar02}. The simplified symmetric states, known as the conventional Dicke states \cite{dicke}, represent the case where it is assumed that the mean separation between the atoms is much less than the wavelength corresponding to the two level transition (which, for the co-propagating off resonant Raman excitation, is $\sim (k_1-k_2)^{-1}$). While this constraint is not necessary for the concept proposed here \cite{shahriar02}, it is easier to describe the process initially under this constraint. The observables computed remain correct when this constraint is not met. Some of these Dicke states are as follows: $|E_0\rangle \equiv |1 1 1 ... 1 \rangle$, $|E_1\rangle \equiv \sum_{i=1}|1 1 ...2_i... 1 \rangle/\sqrt{N}$, $|E_2\rangle \equiv \sum_{i,j\neq i}|1 1 ...2_i...2_j... 1 \rangle /\sqrt{^NC_2}$, $|E_3\rangle \equiv \sum_{i,j,k}|1 1 ...2_i...2_j...2_k... 1 \rangle /\sqrt{^NC_3}$, and $|E_N\rangle \equiv |2 2 2 ... 2 \rangle$ where $^NC_n=N!/n!(N-n)!$. For instance, $|E_2\rangle$ is the Dicke state with two atoms in  $|2\rangle$ and the rest in $|1\rangle$. Any two atoms can be in $|2\rangle$ with equal probability, with $^NC_2=N(N-1)/2$ such possible combinations. 

The Hamiltonian in the basis of the symmetric collective states is $H=\sum^N_{k=0}[ -k\hbar\delta |E_k \rangle \langle E_k|] + \sum^{N-1}_{k=0}[(\hbar\Omega_{k+1}|E_k\rangle \langle E_{k+1}| + H.c.]$ where $\Omega_{k+1}=\sqrt{N-k}\sqrt{k+1}\Omega$ is the Rabi frequency between collective states \cite{shahriar02,dicke}. The states are separated by $\hbar\delta$ in energy and couple at different rates. For instance, $\Omega_1=\Omega_N=\sqrt{N}\Omega$, $\Omega_{2}=\Omega_{N-1}=\sqrt{2(N-1)}\Omega$, etc. The middle states have the strongest coupling rate of $\Omega_{N/2}=N\Omega$ and the end states couple most weakly.  


\begin{figure}
	\includegraphics[width=.48\textwidth]{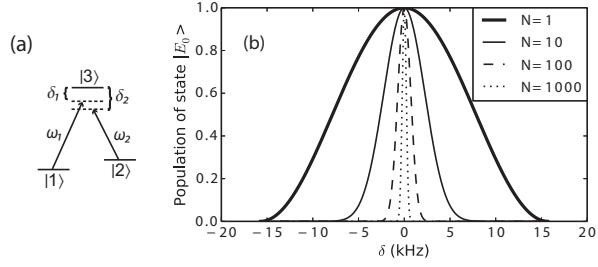}
	\caption{(a) Three level atomic system (b) Population of $|E_N\rangle$ at the end of Ramsey pulse sequence as function of $\delta$.}
	\label{fig:pop}
\end{figure}

The final state of the system at the end of the second $\pi/2$ pulse can be derived by using either the collective state picture or, equivalently, the single atom picture. For a large value of $N$, carrying out the calculation in the collective states basis is numerically cumbersome and analytically intractable. However, we can find the state trivially by using the single atom picture and then determining the coefficients of the collective states by simple projection, given the definition of the $(N+1)$ generalized symmetric collective states. As such, the final state of the system is $|\psi\rangle=\prod_{i=1}^N(W^0_{\pi/2}W^{\delta T_2}_0W^0_{\pi/2})_i|1\rangle_i$. In the basis of the generalized symmetric collective states, this becomes:
\begin{equation}
	|\psi\rangle= (-ie^{i\theta})^N \sum^N_{k=0} \sqrt{^NC_k} (\sin{\theta})^{N-k}(\cos{\theta})^k |\tilde{E}_{k}\rangle
	\label{n_atoms_end}
\end{equation}
The population of the state $|\tilde{E}_N\rangle$ at the end of the separated field experiment is
\begin{equation}
	P_N^C\equiv|\langle \tilde{E}_N|\psi\rangle|^2=(\cos{\theta})^{2N} \label{gg}
\end{equation}
which is simply $(P_2)^N$. This quantity, $P_N^C$, represents the probability of finding the whole system in the state $|E_N\rangle$ whereas $P_2$ represents the probability of finding each atom in state $|2\rangle$. In a conventional experiment, the population of atoms in state $|2\rangle$ is measured, for example, by collecting fluorescence produced by coupling $|2\rangle$ to an auxiliary state. The resulting signal is proportional to $P_2$, independent of the number of atoms. The experiment that we propose, to be described shortly, produces a signal that is proportional to $P_N^C$. When Eq. \eqref{gg} is plotted for various values of $N$ (Fig. \ref{fig:pop} (b)), it is evident that the linewidth of the fringe as a function of $\theta$ decreases as $N$ increases. The value of the linewidth, defined as the full width half maximum (FWHM), is given by $\Gamma(N)=2\arccos{(2^{-1/2N})}$. The derivative of $[\Gamma(1)/\Gamma(N)]^2$ with respect to $N$, for large $N$, approaches the value of $0.8899+O(N^{-3/2})$, which we have verified with a linear fit to $[\Gamma(1)/\Gamma(N)]^2$. To a good approximation, $\Gamma(N)/\Gamma(1)\approx 1/\sqrt{N}$. Noting that $\theta=\delta T_2/2$, $\Gamma(1)\simeq \pi/T_2$ is understood to be the transit time limited linewidth. Then $\Gamma(N)=\Gamma(1)/\sqrt{N}=\pi/(T_2\sqrt{N})$ is a violation of the transit time limit, which is discussed in Section \ref{sec:interpret}, along with the physical interpretation of what occurs in the collective atomic clock system. 

\subsection{Effect of velocity distribution}

A two level atomic system $|\psi\rangle$ interacts with light fields and evolves as $|\psi(t'+t)\rangle = W^{\delta t}_{\Omega t}|\psi (t')\rangle$. The two levels in the proposed scheme are, for example, the hyperfine ground states of an alkali atom such as $^{85}$Rb. After the $\pi/2$-dark-$\pi/2$ sequence, the system is in state $|\psi\rangle=W^{\delta T_2}_{\pi/2}W^{\delta T_2}_0W^{\delta T_2}_{\pi/2}|1\rangle$. Unlike in Section \ref{sec:three}, we here do not make the approximation that $\delta \ll \Omega$. Then the signal we expect to see for a single atom is proportional to  $P_2=|\langle 2|\psi\rangle|^2=|\langle 2|W^{\delta T_2}_{\pi/2}W^{\delta T_2}_0W^{\delta T_2}_{\pi/2}|1\rangle|^2$, and the collective state signal is 
\begin{align}
S_{col}&=\Pi_{i=1}^N{|\langle 2|W^{\delta T_2}_{\pi/2}W^{\delta T_2}_0W^{\delta T_2}_{\pi/2}|1\rangle|^2} \nonumber \\
&=|\langle 2|W^{\delta T_2}_{\pi/2}W^{\delta T_2}_0W^{\delta T_2}_{\pi/2}|1\rangle|^{2N}
\end{align}
We assume that the density of atoms in the trap is fixed at $\rho_{A}=10^9$ mm$^{-3}$, so that the width of the atomic ensemble, which has a Gaussian spatial distribution, varies with the number of atoms. With $N=2\times 10^6$ atoms in the trap, the size of the cigar-shaped ensemble is 1 mm in length in the direction of the Raman beams, and $\sim $ 50 $\mu$m in diameter in the other two directions.

\begin{figure}
	\includegraphics[width=.48\textwidth]{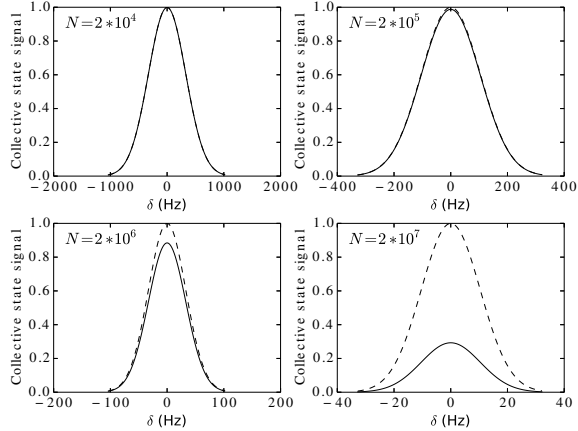}
	\caption{Collective state signal at the end of the Ramsey field experiment for various number of atoms, with parameters $\Omega = 5\cdot 10^5$ s$^{-1}$ and $T_2=3 \cdot 10^{-5}$ s. Plotted are the ideal signal (dashed line), $S_{col}$, and the reduced signal (solid line), $S_{Dop}$, where the effect of Doppler shift is taken into account.}
	\label{fig:doppler}
\end{figure}

When an atom with velocity $v$ interacts with a field with frequency $\omega$ propagating in the direction of the atom, the frequency of the field is shifted by $\delta_D=v\omega/c$. The Maxwell Boltzmann velocity distribution is $\rho_{MB}(v,T) = \sqrt{m_a/(2\pi kT)}exp^{-m_av^2/(2kT)}$ where $m_a$ is the atomic mass and $T$ is the temperature. We assume the temperature to be given by the Doppler cooling limit, so that $T_{MOT} = \Gamma_{Rb}\hbar/(2k) = 138$ $\mu$K for $^{87}$Rb. The average velocity is then $v_{av}\sim 18.3$ cm/s, with a corresponding Doppler shift of $\delta_{D_{av}}=4.18$ Hz. Under these conditions, the signal is
\begin{widetext}
\begin{equation}
S_{Dop}=\Pi_{v'=-5v_{av}}^{5v_{av}} |\langle 2|W^{(\delta+\delta_D(v')) T_2}_{\pi/2}W^{(\delta+\delta_D(v')) T_2}_0W^{(\delta+\delta_D(v')) T_2}_{\pi/2}|1\rangle|^{[2\rho_{MB}(v',T_{MOT})]}
\end{equation}
\end{widetext}
where we take into account velocities that are up to five times the $v_{av}$. Plotted in Fig. \ref{fig:doppler} are the signals $S_{col}$ and $S_{Dop}$ for various $N$ values, with $T_2=3\cdot 10^{-5}$ s and $\Omega=5\cdot 10^6$  s$^{-1}$.  The Doppler effect decreases the overall signal while having virtually no effect on its width. It decreases exponentially as $N$ increases. However, for the given choice of temperature and $N=2\cdot 10^6$, the reduced signal is $S_{Dop} \sim 0.9 S_{col}$. Of course, the signal can be improved if the temperature is reduced below the Doppler cooling limit.

\begin{figure}
	\includegraphics[width=.48\textwidth]{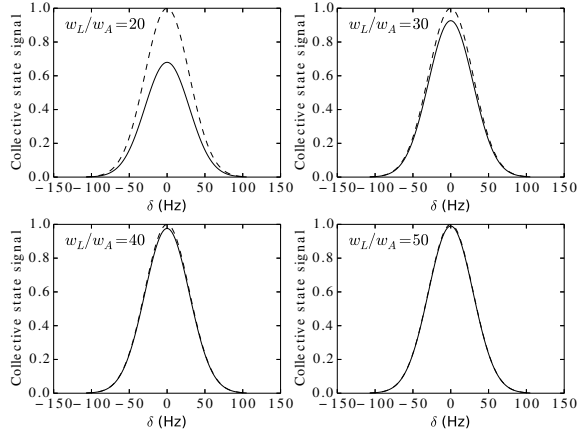}
	\caption{Collective state signal at the end of the Ramsey field experiment for various Gaussian beam widths. $N=2\cdot 10^6$ atoms; $\Omega = 5\cdot 10^5$  s$^{-1}$; $T_2=3 \cdot 10^{-5}$ s. Plotted are the ideal signal (dashed line), $S_{col}$, and the reduced signal (solid line), $S_{\Omega}$, when the Gaussian nature of the beam is taken into account. The plots are for various ratios of the widths of the laser field to the atomic ensemble: $\omega_L/\omega_A$.}
	\label{fig:rabi}
\end{figure}

\subsection{Effect of field inhomogeneity}

Consider next the effect of the inhomogeneity in the laser field amplitude. We assume that the atomic ensemble has a Gaussian spread with a width of $\omega_A$: $\rho_N(\gamma)=\rho_0 e^{-(\gamma^2/\omega_A^2)}$. The width considered in this section is  in the direction perpendicular to the propagation direction of the Raman beams, since the atoms spread in the propagation direction of the beams see the same fields. Each of the two laser fields that produce the Raman-Rabi excitation is also assumed to have a Gaussian profile with a width of $\omega_L>\omega_A$. Since the Raman-Rabi frequency is proportional to the product of the Rabi frequencies for each of these lasers, it follows that the Raman-Rabi frequency is also a Gaussian with a width of $\omega_L$: $\Omega(\gamma)=\Omega_0 e^{-(\gamma^2/\omega_L^2)}$. The peak value of $\Omega$ (i.e., $\Omega_0$) is chosen so that the atoms at the center ($r=0$) experience a perfect $\pi/2$-pulse for an interaction time of $T_1$. Ignoring the effect of the Doppler spread in the velocity, the COSAC signal is then given by 
\begin{equation}
S_{\Omega}=\Pi_{r=-w_{rb}}^{w_{rb}} |\langle 2|W^{\delta T_2}_{\Omega'(r)T_1}W^{\delta T_2}_0W^{\delta T_2}_{\Omega'(r)T_1}|1\rangle|^{2\rho_{N}(r)}
\end{equation}
The signals for various ratios of $w_L/w_A$ are plotted in Fig. \ref{fig:rabi} for $N=2\cdot 10^6$ and density of $\rho_{A}=10^9$ mm$^{-3}$. The signal affected by the inhomogeneous fields can reach the peak value of the ideal signal when $\omega_L/\omega_A=50$. Since $w_A = 50 \mu$m in our system, $w_L=2.5$ mm for the Raman beams is sufficiently large enough to achieve this goal. 
 
\subsection{Effect of spontaneous emission}

In the analysis of the COSAC, we have used a model in which the intermediate state is adiabatically eliminated. However, the actual population of this state is approximately ~$\Omega_1^2/\Delta^2$ with $\Omega_1\sim\Omega_2$. In the time that it takes for a $\pi = \Omega/T_1 \simeq \Omega_1^2/(2\Delta T_1)$ pulse (or two $\pi/2$ pulses) to occur, we can estimate that the number of spontaneous emissions that occur per atom is $(\Omega_1^2/\Delta^2)\Gamma T_1\simeq 2\pi\Gamma/\Delta$. For $\Delta=200 \Gamma$, this, number is about $3\times 10^{-2}$, and increases by a factor of $N$ for an ensemble of $N$ atoms. (Note that there is no enhancement of the rate of spontaneous emission due to superradiant effects, since we are considering a dilute ensemble). As a result, the signal for both the CC and the COSAC would deviate from the ideal one. The actual effect of spontaneous emission on the CC can be taken into account by using the density matrix equation for a three level system. However, in this case, it is not possible to ascribe a well defined quantum state for each atom. This, in turn, makes it impossible to figure out the response of the COSAC, since our analysis for the COSAC is based on using the direct product of the quantum state of each atom. For a large value of $N$, it is virtually impossible to develop a manageable density matrix description of the system directly in terms of the collective states. However, it should be possible to evaluate the results of such a density matrix based model for a small value of $N$ ($<$ 10, for example). In the near future, we will carry out such a calculation and report the findings. 

For the general case of large $N$, one must rely on an experiment (which, in this context, can be viewed as an analog computer for simulating this problem) to determine the degree of degradation expected from residual spontaneous emission. It should be noted that the deleterious effect of spontaneous emission, for both the CC and the COSAC, can be suppressed to a large degree by simply increasing the optical detuning while also increasing the laser power. This is the approach used, for example, in reducing the effect of radiation loss of atoms in a far off resonant trap (FORT).

\subsection{Effect of fluctuation in number of atoms}
For both the CC and the COSAC, the signal is collected multiple times and averaged to increase the signal to noise ratio (SNR). However, the number of atoms can fluctuate from shot to shot. When $N$ fluctuates by $\Delta N$, the  signal in the CC changes by the same amount while the linewidth does not change. It is easy to see this from the classical signal, $S_{CC}=N \cos^2{\theta}$. Changing $N$ by $\Delta N$ will change the signal, but the FWHM, which occurs at $S_{CC}=N/2$, will not change, A more thorough approach for expressing the classical and quantum noise of the CC and the COSAC is covered in Section \ref{sec:performance}-A. In this section, we focus on how the fluctuation in the number of atoms from shot to shot affects the signal of the COSAC.

\begin{figure}
	\includegraphics[width=.48\textwidth]{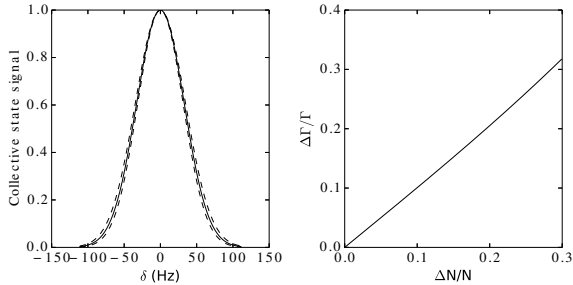}
	\caption{(left) Collective state signal (solid line) at the end of the Ramsey field experiment for $N=2\cdot 10^6$ atoms; $\Omega=5\cdot 10^6$  s$^{-1}$ ; $T=3\cdot 10^{-5}$ s. The dashed curves show the signal for $N+\Delta N$ (narrower) and $N-\Delta N$ (wider), where $\Delta N/N=0.01$.  (right) Plot of $\Delta \Gamma/\Gamma$ as a function of $\Delta N/N$.}
	\label{fig:n_fluct}
\end{figure}

Fig. \ref{fig:n_fluct} (left) is the plot of a collective signal with $N = 2\cdot 10^6$. The dashed red lines represent the case in which $\Delta N/N =0.01$. Increasing the number of atoms by $\Delta N$ decreases the linewidth, and decreasing the number of atoms by $\Delta N$ increases the linewidth. However, the peak of the signal remains at unity. This is in contrast to the results from velocity distribution and field inhomogeneity. We calculate the change in the COSAC linewidth by noting that its FWHM is approximately $\Gamma(N)=\Gamma(1)/\sqrt{N}$. The width of the uncertainty in $\Gamma(N)$, as a result of fluctuation in $N$, is $\Delta\Gamma(N)=\Gamma(1)/\sqrt{N-\Delta N}-\Gamma(1)/\sqrt{N+\Delta N}$, so that the fractional fluctuation is  $\Delta\Gamma(N)/\Gamma(N)=(1-\Delta N/N)^{-1/2}-(1+\Delta N/N)^{-1/2}=\Delta N/N + 0.625 (\Delta N/N)^3+O[(\Delta N/N)^7]$. For small $\Delta N/N$, the fractional change in FWHM is $\Delta \Gamma(N)/\Gamma(N)\simeq \Delta N/N$ to a good approximation. Fig. \ref{fig:n_fluct} (right) shows this correspondence for $N = 2\cdot 10^6$. However, the plot is equivalent for any $N$, since the fractional change in FWHM is only dependent on $\Delta N/N$. 

\section{Experiment and detection scheme for realizing a COSAC}
\label{sec:experiment}

\begin{figure}
	\includegraphics[width=.48\textwidth]{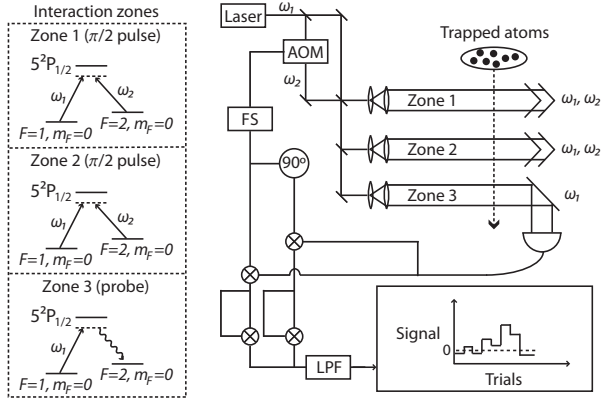}
	\caption{Ramsey fringe experiment for an ensemble of $\Lambda$ type atoms for the detection of collective state $|E_N\rangle$. Atoms are released from the trap, and the experiment is performed while they are free falling inside the vacuum chamber. They interact with two $\pi/2$ Ramsey pulses, which are separated in time by $T_2$, and are probed by a probe. The probe induces a unidirectional Raman transition in the atoms while producing Stokes photons in the direction of the detector, given high enough optical density. The combined signal from the probe and emitted Stokes photons are multiplied with the frequency produced by the FS in such a way that the resulting signal will be proportional to the number of Stokes photons detected. Determining the threshold of the zero emission signal, and counting how many trials result in zero emission, the histogram can be built to produce signal in Fig. \ref{fig:pop}(b).}
	\label{fig:exp}
\end{figure}
\begin{figure}
	\includegraphics[width=.22\textwidth]{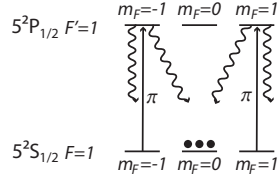}
	\caption{Initialization of the system involves first optically pumping the atoms into $|F=1\rangle$ state by applying a laser field that is resonant with $5^2S_{1/2},|F=2\rangle \to 5^2P_{3/2},|F'=2\rangle$ transition. Afterwards, as is depicted here, a $\pi$ polarized beam that is resonant with $5^2S_{1/2},|F=1\rangle \to 5^2P_{1/2},|F'=1\rangle$ transition is applied. Because the $|F=1,$ $m_F=0\rangle \to |F'=1,$ $m_{F'}=0\rangle$ transition is forbidden for the D1 line, the atoms are eventually pumped into $|F=1,$ $m_F=0\rangle$.}
	\label{fig:opt_pump}
\end{figure}

Before proceeding further, we describe the experimental approach that can be used to measure $P_N^C$, as summarized in Fig. \ref{fig:exp}. For concreteness, and without loss of generality, we consider $^{87}$Rb as the atomic species. By making use of the necessary D2 line transitions, we start by trapping atoms in a magneto-optical trap (MOT), and transferring them into a more localized dipole trap, cooled down to the Doppler cooling limit of $T_D=\hbar\Gamma/(2 k_B)=138$ $\mu$K \cite{foot, ketterle, kowalski, miller}. After capturing about $2\cdot 10^6$ atoms in a cigar shaped cloud with a diameter of $\sim w_A=50 \mu$m and length of 1 mm, the atoms are released and optically pumped into the $|F=1\rangle$ state by applying a beam that is resonant with $5^2S_{1/2},|F=2\rangle \to 5^2P_{3/2},|F'=2\rangle$ transition of rubidium D2 line. Furthermore, a $\pi$ polarized beam that is resonant with $5^2S_{1/2},|F=1\rangle \to 5^2P_{1/2},|F'=1\rangle$ transition of rubidium D1 line is applied, as depicted in Fig. \ref{fig:opt_pump}. Because the $|F=1$, $m_F=0\rangle \to |F'=1$, $m_{F'}=0\rangle$ transition is forbidden for the D1 line, the atoms will finally be pumped into $|F=1$, $m_F=0\rangle$ level. It is possible, with the imperfections that are inadvertently present in the system, that there might be some residual atoms left in $|F=1$, $m_F=-1\rangle$ and $|F=1$, $m_F=1\rangle$. We avoid the detection of these residual atoms by making use of the fact that the Zeeman shifts of levels in $|F=1\rangle$ and $|F=2\rangle$ are in opposite directions, which will be discussed in more detail after we outline the null measurement scheme.
Once the initialization of atoms into $|F=1$, $m_F=0\rangle$ state is complete, a bias magnetic field of $\sim$2 G, generated with a pair of Helmholtz coils, is turned on in the $\hat{z}$ direction. While the atoms are in free fall, we turn on a pair of co-propagating right circularly polarized ($\sigma_+$) Raman beams in the $\hat{z}$ direction. One of these beams is tuned to be $\sim$3.417 GHz red detuned from the $|F=1\rangle \to |F'=1\rangle$ transition (D1 manifold), and the other is tuned to be $\sim$3.417 GHz red detuned from the $|F=2\rangle \to |F'=1\rangle$ transition (D1 manifold). The second Raman beam is generated from the first one via an acousto-optic modulator (AOM), for example. The AOM is driven by a highly stable frequency synthesizer (FS), which is tuned close to $\sim$6.835 GHz corresponding to the frequency difference between the $|F=1\rangle$ and $|F=2\rangle$ states in the $5^2S_{1/2}$ manifold.
\begin{figure}
	\includegraphics[width=.3\textwidth]{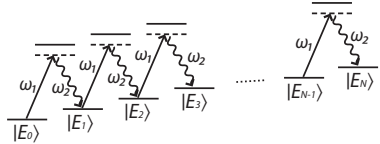}
	\caption{In the detection zone, we probe the population of state $|E_N\rangle$ by applying field $\omega_1$ and detecting Stokes photons produced during the Raman transition. In the bad cavity limit, the atomic system will not reabsorb the photon that has been emitted during the Raman process, such that the transition from $|E_k\rangle$ to $|E_{k+1}\rangle$ will occur, but not vice versa.}
	\label{fig:int}
\end{figure}

These beams excite off-resonant Raman transitions between $|F=1$, $m_F=m\rangle$ and $|F=2$, $m_F=m\rangle$ levels, for $m=1,0,-1$. Since the system is initialized in $|F=1$, $m_F=0\rangle$, the $\sigma_+$ Raman transitions through the excited states $|F'=1$, $m_{F'}=1\rangle$ and $|F'=2$, $m_{F'}=1\rangle$ couple the initial state to $|F=2$, $m_F=0\rangle$. Hence, the energy levels $|1\rangle$ and $|2\rangle$ from the previously discussed $\Lambda$ scheme correspond to hyperfine ground states $|F=1$, $m_F=0\rangle$ and $|F=2$, $m_F=0\rangle$, respectively. The resulting four level system, with the two excited states, can be reduced to a two level system in the same manner as the $\Lambda$ system by adiabatically eliminating the excited states together. The resulting two level system has a coupling rate that is the sum of the two Raman Rabi frequencies, one involving the $|F'=1$, $m_{F'}=1\rangle$ state, and the other involving the $|F'=2$, $m_{F'}=1\rangle$ state. The laser power at $\omega_1$ and $\omega_2$ are adjusted to ensure that the light shifts of levels $|1\rangle$ and $|2\rangle$ are matched. 

In the first interaction zone, the co-propagating Raman beams interact with the atomic ensemble for a duration of $\Omega T_1=\pi/2$. After waiting for a time $T_2$, chosen such that $T_2\gg T_1$, we pulse the Raman beams again, in place, to interact with the atomic ensemble for another duration $\Omega T_1=\pi/2$. The Raman beams can be pulsed in place as long as the width of the beams is much larger than that of the free-falling, thermally expanding atomic cloud. 

After these excitations, we probe the population in one of the collective states, $|E_N\rangle$ , where all the individual atoms are in state $|2\rangle$, by a method of zero photon detection. For illustrative purposes, let us consider first a situation where the atomic ensemble is contained in a single mode cavity with mode volume $V$, cavity decay rate $\gamma_c$,  and wavevector $k_2=\omega_2/c$. The cavity is coupled to the atomic transition $|2\rangle\rightarrow|3\rangle$ with coupling rate $g_c=|e \langle r \rangle|E/\hbar$, where $|e \langle r \rangle|$ is the dipole moment of the atom and the field of the cavity is $E=\sqrt{2\hbar\omega_2/(\epsilon_0 V)}$. If we then send a probe beam, an off-resonant classical laser pulse with frequency $\omega_1$, the presence of the cavity will allow Raman transitions to occur between the collective states $|E_k\rangle$ and $|E_{k+1}\rangle$ with the coupling rates $\Omega'_{k+1}=\sqrt{N-k}\sqrt{k+1}\Omega'$ where $\Omega'=\Omega_1g_c/2\Delta$. The schematic of the interaction is shown in Fig. \ref{fig:int}. 

In the bad cavity limit where $\gamma_c\gg \sqrt{N}\Omega'$, the Raman transitions will still occur. However, the atomic system will not reabsorb the photon that has been emitted during the process, such that the transition from $|E_k\rangle$ to $|E_{k+1}\rangle$ will occur, but not vice versa. The electric field of such a photon is $E=\sqrt{2\hbar\omega_2/(\epsilon_0 Ac\tau)}$, where $A$ is the cross sectional area of the atomic ensemble, $c$ is the speed of light, and $\tau$ is the duration of the photon.  This limit applies in our case, which has no cavity. In this limit, the stimulated Raman scattering is an irreversible process that can be modeled as a decay with an effective decay rate that is singular to each $|E_j\rangle$ state. The decay rate from state $|E_1\rangle$ is ${\gamma }_0=4NL{{\left| g_c\Omega_1  \right|}^{2}}/({{\Delta }^{2}}c)=N{{\gamma }_{sa}}$ where $\gamma_{sa}=16L\Omega'^2/c$ \cite{dlcz} is the decay rate for a single atom. The value of $g_c$ is given by $|e\langle r \rangle|\cdot E$. The effective decay rates for the other states can be calculated following the same logic as $\gamma_j=(j+1)(N-j)\gamma_{sa}$. 

When photons are scattered through stimulated Raman scattering in the detection process, the resonant optical density (OD) determines the degree to which the emission occurs in the direction of propagation of the probe beam \cite{dlcz}. Specifically, the fraction of photons that are not emitted in the direction of the probe is give by $1/OD$.  Thus, $(1-1/OD)$ determines the effective collection efficiency of the detection process. The OD depends on the density of atoms $n$, the diameter of the atomic ensemble $w_A$, and and the resonant scattering cross section $\sigma \simeq (\lambda/2)^2$, as $\rho=\sigma n L$. For the rubidium-87 D1 line wavelength, $\lambda\sim 795$ nm, and a cigar shaped trap with $N=2\cdot 10^6$ atoms, a diameter of 50 $\mu$m, and a length of 1 mm, we find that the resonant optical density is $\rho\sim 300$. The beam consisting of the probe and the emitted photons is sent to a high speed detector, which produces a dc voltage as well as a signal at the beat frequency of $\sim 6.835$ GHz. The phase of this beat frequency signal is unknown. As such, the total signal is sent in two different paths, one to be multiplied by the FS signal and another to be multiplied by the FS signal shifted in phase by 90 degrees. Each of these signals is squared, then combined and sent through a low pass filter (LPF) to extract the dc voltage that is proportional to the number of scattered photons. A voltage reading above a predetermined threshold value will indicate the presence of emitted photons during the interrogation period. The interrogation period is set to $\gamma_0 T = 10$ where $\gamma_0=\gamma_{N-1}=N\gamma_{sa}$ is the slowest decay rate, to ensure that even the longest lived state has a chance to decay almost completely. If no photon emission occurs and the voltage reads below the threshold, this indicates that the atoms are all in $|2\rangle$ and the collective state of the system is $|E_N\rangle$. For any other collective state, at least one photon will be emitted. For a given value of $\delta$, this process is repeated $m$ times (where the choice of $m$ would depend on the temporal granularity of interest). The fraction of events corresponding to detection of no photons would represent the signal for this value of $\delta$. The process is now repeated for a different value of $\delta$, thus enabling one to produce the clock signal as a function of $\delta$. Usual techniques of modulating the detuning and demodulating the signal can be used to produce the error signal for stabilizing the FS, thus realizing the COSAC.

As noted earlier, it is possible that a small fraction of the detected signal might be due to the residual atoms that were not optically pumped to $|F=1,$ $m_F=0\rangle$ initially. The $\sigma_+$ polarized Raman probe is applied to $|F=1\rangle$ level, and the residual atoms in $|F=1,$ $m_F=-1\rangle$ and $|F=1,$ $m_F=1\rangle$ can also see the excitation. However, the bias magnetic field of 2 Gauss lifts the degeneracy of the energy levels. Moreover, since $g_F=-1/2$ for $|F=1\rangle$ and $g_F=1/2$ for $|F=2\rangle$, the energy levels shift in opposite directions such that the Raman signals for the transitions involving $m_F=-1$ and $m_F=1$ are detuned from resonance. Each will be shifted by $\delta_z=-m(g_{F=2}-g_{F=1})\mu_B B/\hbar=-1.4$ [MHz/Gauss] $\cdot m_F B$ where $B=2 $ Gauss. Therefore, these transitions will not be a part of the detection, which only involves looking at 6.835 GHz beat frequency between the probe and the spontaneously generated photon.

In the particular implementation of the COSAC considered here, we have used off-resonant Raman transition. However, effects such as residual light shifts can limit the stability of such a clock. The ground states can also be coupled directly by using a microwave pulse, which has the advantage of being free from differential light shifts. Thus, the COSAC can also be realized by using a traveling wave microwave pulse sequence for the separated Ramsey field experiment \cite{hemmer}, as long as the detection pulse remains the same. Since the Hamiltonian for light-shift balanced off-resonant Raman excitation, with the excited state eliminated adiabatically, is formally identical to that of microwave excitation \cite{shahriar00}, the basic behavior of the COSAC would be identical for a microwave version. 

	
\section{Performance of the COSAC compared to that of the CC}
\label{sec:performance}

In order to compare the performance of the COSAC to that of the comparable CC, we examine the stability of the clocks by investigating the fluctuation that has both quantum mechanical and classical components, or  $\delta f|_{total}=(\Delta S_{QM}+\Delta S_{class})/(\partial S / \partial f)$, where $S(f)$ is the signal and $f$ is the detuning of the clock away from its center value. Because the signal depends on the frequency, the fluctuations in a clock are not necessarily constant, and there is not a single value of the SNR to compare unless we compare the two clocks at a particular value of the frequency. Instead, the fluctuations must be compared as a function of $f$ for completeness. In this section, we discuss the quantum fluctuation due to quantum projection noise, $\Delta P = \sqrt{P(1-P)}$ \cite{wineland01}, where $P$ is the population of the state to be measured, the classical noise in the long term regime, and the effects of detector efficiency and the collection efficiency. The ratios of the frequency fluctuations in the CC to the frequency fluctuations in the COSAC show that the two clocks perform comparably around the signal at $f=0$ if the clocks have perfect collection efficiency. However, the traditional fluorescence detection based clock suffers from collection efficiency issues that the collective clock is immune to. For the CC, a resonant beam probes the clock state, generating spontaneously emitted photons. The collection efficiency of such a system is limited by the solid angle of the detection system. On the other hand, the COSAC collects the fluorescence of photons through coherent Raman scattering, which enables large collection efficiency that can be close to unity for sufficiently high resonant optical density (as noted earlier). As such, for the same number of atoms detected per unit time, the COSAC is expected to perform better than the fluorescence detection based CC by as much as a factor of 10. This is discussed in greater detail in subsection C of this section.

\subsection{Effects of quantum and classical noise}
In order for the COSAC to be useful, it must perform at least as well as, or better than, the CC, and for that, we must compare the two clocks' stability in the short term and the long term regimes. The stability of a clock can be measured by investigating the frequency fluctuation that has both quantum mechanical and classical components. Before comparing the stabilities of the COSAC and the CC, it is instructive first to review briefly the stability of a CC. 

For concreteness, we consider an off-resonant Raman-Ramsey clock as the CC. The population of the detected state $|2\rangle$ at the end of the second pulse is given by $P_2=\cos^2{(fT_2/2)}$, where $T_2$ is the separation period of the two $\pi/2$-pulses and $f$ is the deviation of the clock frequency away from its ideal value, expressed in radial units (i.e. rad/s rather than Hz). The signal is detected by probing the desired state for a duration of time. If $\tilde{N}$ is the number of atoms per unit time and $\tau$ is the interrogation period, the net signal is $S_{sa}=\tilde{N}\tau P_2=\tilde{N}\tau\cos^2(fT_2/2)$. For the sake of comparison, we allow the number of atoms per trial in the COSAC signal, $N$, multiplied by the number of trials, $m$, to equal $\tilde{N}\tau$. Therefore, we can write $S_{sa}=mN\cos^2(fT_2/2)$. The quantum mechanical variance of this quantity is $\Delta S_{QM,sa}=(\sqrt{mN}/2)\sin({fT_2})$, where the derivation is made by noting that the fluctuations in $mN$ is $\sqrt{mN}$ \cite{wineland01}, and the projection noise in a single two level atomic system is $\Delta P_{2}=\sqrt{P_2(1-P_2)}$ \cite{wineland01}. (It should be noted that the fluctuation in $mN$ is also a manifestation of this projection noise, as discussed in detail in \cite{wineland01}.) When the probability of finding the population in this state is unity or nil, the projection noise vanishes; on the other hand, it is largest at $P_2=1/2$. Calculating the slope from the signal, we find that 
$\partial S_{sa}/\partial f=-[mN/(2\gamma_{sa})]\sin({f T_2})$,
where $\gamma_{sa}=1/T_2$ is the linewidth. 

Assuming perfect quantum efficiency for the detection process, the frequency fluctuation can be written as $\delta f|_{total}=|(\Delta S_{QM}+\Delta S_{class})/(\partial S / \partial f)|$, which can be regarded as noise ($\Delta S$), both quantum and classical, over the Spectral Variation of Signal ($\partial S / \partial f$), or SVS. In what follows, we consider first the effect of quantum noise only. Thus, the quantum frequency fluctuation (QFF) for a CC can be expressed as
\begin{equation}
\partial f_{QM,CC}\equiv \left| \frac{\Delta S_{QM, sa}}{(\partial S_{sa}/\partial f)}\right|=\frac{\gamma_{sa}}{\sqrt{mN}}
\label{cc_qff}
\end{equation}
It should be noted that while both $\Delta S_{QM}$ and $(\partial S/\partial f)$ depend on $f$, their ratio is a constant, which is merely an accident due to the fact that the signal is cosinusoidal. However, this accidental cancellation has led to an apparently simple perception of the QFF as being simply the ratio of the linewidth ($\gamma_{sa}$) to the SNR, where the SNR is understood to be $\sqrt{mN}$. This expression for the SNR, in turn, follows from thinking about the signal as being $S'=mN$ and noise $N'$ as being $\sqrt{mN}$, so that SNR$\equiv S'/N'=\sqrt{mN}$. However, it should be clear from the discussion above that the signal is not given by $mN$, and noise is not given by $\sqrt{mN}$; rather, they both depend on $f$. 

In cases where frequency fluctuation is not a constant (as will be the case for the COSAC), we can no longer measure the stability of the clock in terms of a constant $\gamma/$SNR. Instead, it is necessary to carry out the full calculation of the frequency fluctuation as a function of frequency. Thus, we will adopt the convention that the net frequency fluctuation, $\delta f$, should be thought of as the \textit{ratio of the noise to the SVS}. This approach should be adopted universally for all metrological devices. Of course, for devices where the relevant quantity is not the frequency, the definition should be adapted accordingly. For example, in an interferometer that measures phase, the relevant quantity can be expressed as follows: net phase fluctuation is the  \textit{ratio of the noise to the Angular Variation of Signal (AVS)}. 

Following this convention, we can now examine the net frequency fluctuation of the COSAC and compare it to that of the CC. We will first compare their quantum fluctuations, which is relevant in the short term regime, and then the classical fluctuations, which dominates the long term regime. The collective state signal for $m$ trials is $S_{col}=mP^C_N=m\cos^{2N}{(fT_2/2)}$ and the projection noise is $\Delta P^C_N=\sqrt{P^C_N(1-P^C_N)}$ for a single trial and $\Delta P^C_N=\sqrt{m}\sqrt{P^C_N(1-P^C_N)}$ for $m$ trials, so that the total quantum mechanical noise in the signal is 
\begin{equation}
\Delta S_{QM,col}=\sqrt{m}\cos^{N}{(fT_2/2)}\sqrt{1-\cos^{2N}{(fT_2/2)}}
\end{equation}
and the SVS is 
\begin{equation}
\partial S_{col}/\partial f=-(mN/\gamma_{sa}) \sin{(fT_2/2)} \cos^{2N-1}{(fT_2/2)}
\end{equation}
Therefore, the frequency fluctuation in the COSAC due solely to quantum noise can be expressed as:
\begin{equation}
\delta f_{QM,COSAC}=
\left|\frac{\gamma_{sa}}{N\sqrt{m}}\sqrt{\frac{1-P^C_N}{P^C_N}}\cot\left({\frac{fT_2}{2}}\right)\right|
\label{COSAC_qff}
\end{equation}
where $P_N^C$ is a function of $f$. Thus, unlike in the case of the CC, the frequency fluctuation is not a constant, and depends strongly on $f$. 

We consider first the limiting case of $f \rightarrow 0$. Using Taylor expansion, it is easy to see that 
\begin{equation}
\delta f_{QM,COSAC}\simeq
\frac{\gamma_{sa}}{\sqrt{mN}}
\end{equation}
which is the same as that of the CC, given in Eq. \eqref{cc_qff}. This can be understood physically by noting that while the fringe width becomes much narrower for the COSAC, the SNR also decreases due to the fact that a single observation is made for all N atoms in a given trial. 

\begin{figure}
	\includegraphics[width=.48\textwidth]{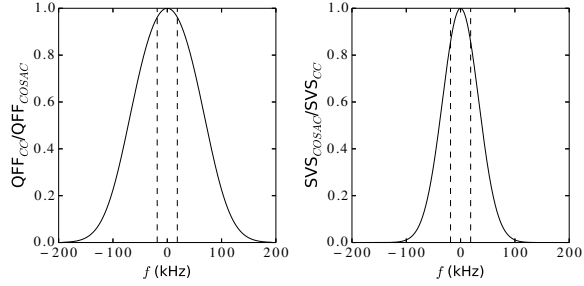}
	\caption{(left) Ratio of the QFF in the CC to the QFF in the COSAC, for $T_2=3\cdot 10^{-5}$ s, $m=1000$ and $N=2\cdot 10^6$. It should be noted that the fluctuation in the CC is independent of $f$ while that of the COSAC varies significantly with $f$. (right) Ratio of the SVS of the COSAC to the SVS of the CC for $T_2=3\cdot 10^{-5}$ s, $m=1000$ and $N=2\cdot 10^6$. The dashed vertical lines in the plots show where the $FWHM_{col}$ are. }
	\label{fig:snr_comp}
\end{figure}

The QFF for the COSAC, given in Eq. \eqref{COSAC_qff}, is smallest as $f\rightarrow 0$ and increases as $f$ moves away from resonance. The ratio of the QFF for the CC, given in Eq. \eqref{cc_qff}, to that of the COSAC, given in Eq. \eqref{COSAC_qff}, is plotted as a function of $f$ in Fig. \ref{fig:snr_comp} (left) for $T_2=10^{-4}$ s, $m=1000$ and $N=2\cdot 10^6$. Here, the vertical bars indicate the FWHM of the COSAC signal. It is clear from this plot that the QFF for the COSAC increases significantly as we move away from resonance. However, since a servo will keep the value of $f$ confined to be close to zero, the frequency stability of the COSAC, under quantum noise limited operation, should be very close to that of the CC, assuming that all the other factors remain the same. 

The classical frequency fluctuation (CFF), $\partial f|_{class}=\Delta S_{class}/(\partial S/\partial f)$, is the limiting factor in the long term stability. While the quantum fluctuation is dominated by quantum projection noise, the classical noise is dominated by noise in the electronics employed to generate the clock signal. Since the pieces of equipment used in the development of both the COSAC and CC suffer from similar noise issues, the variance $\Delta S$ is expected to be of the same order of magnitude for both clocks. On the other hand, the SVS, $(\partial S/\partial f)$, is not the same, as was shown previously. The ratio of the SVS of the COSAC to the SVS of the CC is
\begin{equation}
\frac{\partial S_{col}/\partial f}{\partial S_{sa}/\partial f}=\frac{\cos^{2N}{\left(\frac{fT_2}{2}\right)}}{\cos^{2}{\left(\frac{fT_2}{2}\right)}}=\frac{P^C_N}{P_2}
\end{equation}
and is plotted in Fig. \ref{fig:snr_comp} (right). With $\Delta S_{class, col}\sim\Delta S_{class, sa}$, the ratio of the CFF of the COSAC to the CFF of the CC can be written
\begin{equation}
\frac{\delta f_{class,COSAC}}{\delta f_{class,CC}}\simeq\frac{\cos^{2}{\left(\frac{fT_2}{2}\right)}}{\cos^{2N}{\left(\frac{fT_2}{2}\right)}}
\label{ratio_cff}
\end{equation}
Similar to the ratio of the two clocks in QFF, Eq. \eqref{ratio_cff} is smallest as $f\rightarrow 0$ and increases as $f$ moves away from resonance. Thus, with respect to both quantum and classical sources of noise, the COSAC must be operated near $f\simeq 0$ for optimal performance.

We have investigated the effects of quantum and classical noise by deriving the expression for fluctuation in frequency. However, as was shown in the first section, the signal is also a function of other experimental variables; and in general, the fluctuations in any of these can be expressed as
\begin{equation}
\partial A\equiv \left| \frac{\Delta S_{QM}(A)+\Delta S_{class}(A)}{\partial S(A)/\partial A}\right|
\label{fluctuation}
\end{equation}
where $A$ is the variable whose fluctuation is of interest, and the signal $S$ is expressed in terms of $A$.

\subsection{Effect of detector efficiency}
We recall briefly that in the COSAC detection scheme, a laser with a frequency corresponding to one leg of the Raman transition interacts with the atoms, which are in the quantum state $|\psi\rangle =c_N|E_N\rangle +\sum\nolimits_{j=0}^{N-1}{c_j| E_j\rangle}$. Interaction between this field, the atoms, and the free space vacuum modes on the other leg would lead to production of photons unless $c_N=1$ and $c_j=0$ for all $j$. These photons are detected using a heterodyning technique, as described previously. The voltage output of the heterodyning system is proportional to the amplitude of the electric field corresponding to the photons. 

In general, one or more photons are produced as $|E_j\rangle$ decays to $|E_{j+1}\rangle$ and subsequent states. The time needed for these photons to be produced depends on the vacuum and probe field induced Raman transition rates between $|E_j\rangle$ and $|E_{j+1}\rangle$. If one assumes perfect efficiency for detecting each of these photons, and waits for a time long compared to the inverse of the weakest of these transition rates, then the detection of no photons implies that the system is in state $|E_N\rangle$. In practice, we can choose a small threshold voltage at the output of the heterodyning system as an indicator of null detection. Thus, any signal below this threshold would be viewed as detection of the quantum system in the $|E_N\rangle$ state, and all signals above this threshold would be discarded. The number of events below this threshold for $m$ trials carried out with all the parameters of the experiment unchanged, is the derived signal for the COSAC. After collecting data for all the values of detuning that is of interest, the result would ideally yield the plot of the COSAC signal  $S_{col}=|c_N|^2$, averaged over $m$ trials. However, with a fractional detector efficiency and finite detection period, the signal would deviate from the ideal result.

Consider first the effect of the detection period. Given the decay rate of the off-resonant Raman process, $\gamma_j=(j+1)(N-j)\gamma_{sa}$ as described previously, the probability that $|E_j\rangle$ will produce zero photons during the measurement period $\tau$ is $P_{0,j}=e^{-\gamma_j\tau}$. Thus, the total probability of zero photon emission (which should vanish ideally for any $c_j\neq 0$) is given by $P_0=\sum\nolimits_{j=0}^{N-1}{{{\left| {{c}_{j}} \right|}^{2}}}{{e}^{-{{\gamma }_{j}}\tau}}$. The collective state signal, $S_{col}$, is the total probability of finding zero photons during $\tau$, and can be expressed as $S_{col}={{\left| c_N \right|}^{2}}+\sum\nolimits_{j=0}^{N-1}{{{\left| {{c}_{j}} \right|}^{2}} e^{-\gamma_j\tau}}$. Noting that $\gamma_N=0$, we can rewrite this compactly as $S_{col}=\sum\nolimits_{j=0}^{N}{{{\left| {{c}_{j}} \right|}^{2}} e^{-\gamma_j\tau}}$. The lower and upper bounds of $S_{col}$ can be established by considering the strongest and the weakest effective decay rates. The strongest decay rate occurs for the middle state, 
$\gamma_{N/2}=(N/2)(N/2+1)\approx (N^2/4)\gamma_{sa}$, where $N \gg 1$ approximation has been made. With the substitution of the largest decay rate for each $|E_j\rangle$ into the equation for $S_{col}$, the lower bound is set by 
\begin{equation}
S_{LB}={{\left| {{c}_{N}} \right|}^{2}}+\left( 1-{{\left| {{c}_{N}} \right|}^{2}} \right){{e}^{-\frac{{{N}^{2}}}{4}{{\gamma }_{sa}}\tau}}
\label{s_lb_1}
\end{equation}
Likewise, with the substitution of the weakest decay rate for each $|E_j\rangle$, $\gamma_0=\gamma_{N-1}=N\gamma_{sa}$, into $S_{col}$, the upper bound is set by
\begin{equation}
S_{UB}={{\left| {{c}_{N}} \right|}^{2}}+\left( 1-{{\left| {{c}_{N}} \right|}^{2}} \right){{e}^{-N{{\gamma }_{sa}}\tau}}
\label{s_ub_1}
\end{equation}
The signal produced in time $\tau$ will then lie somewhere between the lower and the upper bounds.  

Consider next the effect of non-ideal detection efficiency of the heterodyning scheme. To be concrete, let us define as $\eta$ the efficiency of detecting a single photon. In practice, this parameter will depend on a combination of factors, including the quantum efficiency of the high-speed photodetector and the overlap between the probe laser mode and the mode of the emitted photon, as well as the resonant optical depth of the ensemble, as discussed earlier. For the COSAC, it should be noted that we are interested in knowing only whether one or more photons have been detected, and not in the actual number of photons. When more photons are emitted, the detector will have a better chance of observing a non-zero signal, and hence distinguish zero photon emission from the rest with more certainty. For example, if three photons are emitted during the interrogation time, then four different outcomes are possible: 
\begin{itemize}
\item	All three photons are detected, with probability  $\eta^3$;
\item	Two of the photons are detected, with probability  $\eta^2(1-\eta)$; this can occur for any two of the photons, so the multiplicity is 3;
\item	One photon is detected, with probability $\eta(1-\eta)^2$ and multiplicity of 3. 
\item	No photons are detected, with probability $\epsilon^3$ where $\epsilon\equiv 1-\eta$
\end{itemize}
The sum of these probabilities is 1. The probability that at least 1 photon is detected is thus $(1-\epsilon^3)$. For any state $j\neq N$, the probability of detecting at least 1 photon is therefore $(1-\epsilon^{N-j})$.

Moreover, we must also consider how the effective detection efficiency is influenced by the fact that the collective states decay at different rates. Specifically, the $j$th level for $j<N$ might produce $N-j$ photons, $N-j-1$ photons, down to no photons, depending on the length of the measurement time and the effective decay rate. If the system is in the state $|E_{N-3}\rangle $, for example, it can produce up to 3 photons but with probabilities that change over the course of the detection period. For a given time $\tau$, $|E_{N-3}\rangle $ evolves into a sum of the states 
$|E_{N-3}\rangle \to \sum\nolimits_{k=N-3}^{N}{a_{jk}(\tau)|E_k\rangle}$, where the coefficient $a_{jk}(\tau)$ depends on the effective decay rate that is specific to each state, and changes as the states evolve in time. The detector efficiency can be inserted to show the true probability of detecting a non-zero signal, keeping in mind that no photon is produced if the ensemble remains in state $|E_{N-3}\rangle $, 1 photon is produced via evolution of the ensemble to state $|E_{N-2}\rangle $, and so on. Then the probability of at least one photon being produced during a period of $\tau$ is
\begin{equation}
P_{N-3}=\sum\limits_{k=N-3}^{N}{\left( 1-{{\varepsilon }^{k-N+3}} \right){{\left| {{\alpha_{jk}(\tau)}} \right|}^{2}}}
\label{prob_3}
\end{equation}
Thus, the total probability of detecting at least one photon is:
\begin{equation}
P=\sum\limits_{j=0}^{N-1}{{{\left| {c_j} \right|}^{2}}}\sum\limits_{k=j}^{N}{\left( 1-{{\varepsilon }^{k-j}} \right){{\left| {{\alpha }_{jk}(\tau)} \right|}^{2}}}
\end{equation}
The probability of seeing no photon is 
\begin{equation}
S_{col}=1-P=1-\sum\limits_{j=0}^{N-1}{{{\left| {c_j} \right|}^{2}}}\sum\limits_{k=j}^{N}{\left( 1-{{\varepsilon }^{k-j}} \right){{\left| {{\alpha }_{jk}(\tau)} \right|}^{2}}}
\end{equation}

The numerical analysis for a large number of atoms is tedious and scales as at least $(N-1)!$ for the COSAC. However, we can take the worst case scenario to serve as the upper bound for the signal. The worst case occurs when only a single photon is produced as a result of $|E_j\rangle$ decaying to only the $|E_{j+1}\rangle$ state, so that the index of the second summation stops at $k=j+1$. In this case, we can write $|a_{j,j+1}(\tau)|=(1-e^{-\gamma_j\tau})$ and the signal becomes
\begin{equation}
S_{col}={{\left| {{c}_{N}} \right|}^{2}}+\varepsilon \left( 1-{{\left| {{c}_{N}} \right|}^{2}} \right)+\eta \sum\limits_{j=0}^{N-1}{{{\left| {{c}_{j}} \right|}^{2}}{{e}^{-\gamma_j\tau}}}
\label{s_worst}
\end{equation}
Now, using the approach we employed in arriving at equations Eq. \eqref{s_lb_1} and Eq. \eqref{s_ub_1}, we now consider the strongest and the weakest decay rates for single photon production to arrive at the lower and upper bounds of the zero photon count signal:
\begin{align}
S_{LB}&=1-\eta \left( 1-{{\left| {c_N} \right|}^{2}} \right)\left( 1-{{e}^{-\frac{{{N}^{2}}}{4}{{\gamma }_{sa}}\tau}} \right) \\
S_{UB}&=1-\eta \left( 1-{{\left| {c_N} \right|}^{2}} \right)\left( 1-{{e}^{-N{{\gamma }_{sa}}\tau}} \right)
\end{align}

\begin{figure}
	\includegraphics[width=.48\textwidth]{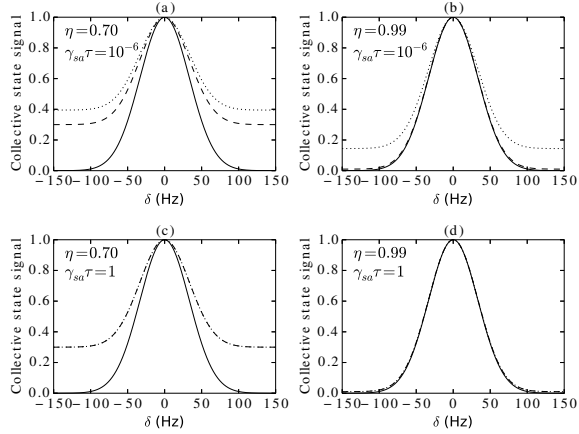}
	\caption{Plot of the ideal signal (solid line), the upper bound (dotted line), and the lower bound (dashed line) for $N=2\cdot 10^6$, $T_2=3\cdot 10^{-4}$ s, and $\gamma_{sa}=10^4$ s$^{-1}$. Note that in (c) and (d), the upper and lower bounds are virtually indistinguishable.}
	\label{fig:error_plot}
\end{figure}

Plots in Fig. \ref{fig:error_plot}  are of the ideal signal (under infinite detection time and $\eta=1$), the lower bound, and the upper bound for various values of $\tau$ and $\eta$ for $N=2\cdot 10^6$, $T_2=3\cdot 10^{-5}$ s, and $\gamma_{sa}=10^4$ s$^{-1}$. As can be seen, the detector efficiency and measurement time do not affect the peak value of the amplitude. As the signal trails off for non-zero detuning, however, the difference increases. The decrease in $\eta$ affects both $S_{UB}$ and $S_{LB}$ similarly, whereas the effect of the decrease in $\tau$ is more evident in $S_{UB}$. With the given parameters, the interrogation period of $\tau=10^{-4}$ s and detector efficiency of $\eta=0.99$ yields almost ideal signal. A somewhat lower value of $\eta$ (e.g. 0.70) still yields a signal that is nearly ideal near zero detuning, which is the desired operating regime for the COSAC, as pointed out earlier. 

If we set $\gamma_{sa}\tau=1$, the signal depends on $\eta$ as
\begin{equation}
S_{col}\simeq 1-\eta\left[1-\cos^{2N}{(fT_2/2)}\right]
\label{s_det}
\end{equation}
for large $N$ and $m=1$. Hence, we can calculate the QFF for the COSAC to see how it depends on the detector efficiency, and how it compares to the CC. For the CC, it is straightforward to show that with $S_{sa}=\eta N \cos^2{(fT_2/2)}$, the quantum mechanical noise in the signal is $\Delta S_{sa}=\sqrt{\eta N}\cos{(fT_2/2)}\sin{(fT_2/2)}$ and the SVS is $|\partial S_{sa}/\partial \delta| =(\eta N/\gamma_{sa})\cos{(fT_2/2)}\sin{(fT_2/2)}$, so that the QFF is $\delta f_{QM,CC}=\gamma_{sa}/\sqrt{\eta N}$. It is also straightforward to calculate the QFF for the COSAC. The total quantum mechanical noise in the COSAC signal in Eq. \eqref{s_det} is:
\begin{equation}
\Delta S_{QM,col}=\sqrt{\eta}\cos^{N}{(fT_2/2)}\sqrt{1-\cos^{2N}{(fT_2/2)}}
\end{equation}
and the SVS is 
\begin{equation}
\partial S_{col}/\partial f=-(\eta N/\gamma_{sa}) \sin{(fT_2/2)} \cos^{2N-1}{(fT_2/2)}
\end{equation}
Thus, the QFF in the COSAC is:
\begin{equation}
\delta f_{QM,COSAC}=
\left|\frac{\gamma_{sa}}{N\sqrt{\eta}}\sqrt{\frac{1-P^C_N}{P^C_N}}\cot\left({\frac{fT_2}{2}}\right)\right|
\end{equation}
which approaches $\gamma_{sa}/\sqrt{\eta N}$ as $f\rightarrow 0$. Assuming that the detector efficiencies of the COSAC and the CC can be essentially the same, they do not affect the ratio of the two QFFs. 

\subsection{Effect of collection efficiency}
We consider next the effect of the collection efficiency, $\beta$. The signal, for both the COSAC and CC, is directly proportional to $\beta$. Thus, it is easy to see, using Eqs. \eqref{cc_qff} and \eqref{COSAC_qff}, that
\begin{align}
\zeta &\equiv\frac{\delta f_{QM,COSAC}}{\delta f_{QM,CC}} \nonumber
\\ 
&=
\left[\frac{1}{\sqrt{N}}\sqrt{\frac{1-P^C_N}{P^C_N}}\cot\left({\frac{fT_2}{2}}\right)\right]\sqrt{\frac{\beta_{CC}}{\beta_{COSAC}}}
\label{ratio_qff}
\end{align}
where $\beta_{CC}$ ($\beta_{COSAC}$) is the collection efficiency of the CC (COSAC). 

As noted above, the quantity written in the square bracket in Eq. \eqref{ratio_qff} approaches unity as $f\rightarrow 0$. Thus, in this limit, we see that the ratio of the QFF for the COSAC to that of the CC would depend on the ratio of the collection efficiencies of the detection processes. As discussed previously, for a high enough resonant optical density ($10^3$ in the example we are considering) the coherent stimulated Raman scattering based detection method used for the COSAC process has a collection efficiency that is close to unity, or $\beta_{COSAC}\simeq 1$. As for the CC, the fluorescence is typically collected from the spontaneous emission process, which emits photons in a dipolar radiation pattern. We can estimate typical values of $\beta_{CC}$ by considering, for example, a CC that makes use of cold atoms released from a MOT. For a lens placed at a distance of 5 cm, with a diameter of 2.5 cm, ignoring the dipolar pattern of radiation for simplicity, and assuming it to be uniform in all directions, this system yields a value of $\beta_{CC}\simeq r^2/(4d^2)=1/16$ corresponding to $\zeta \sim 0.25$. In a typical CC, various geometric constraints make it difficult to achieve a value of $\beta_{CC}$ much larger than this. In fact, in cases where the total volume occupied by the CC has to be constrained in order to meet the user requirements, the value of $\beta_{CC}$ is typically 1$\%$, which would correspond to $\zeta\sim 0.1$. Thus, the near unity collection efficiency of the COSAC can lead to an improvement of the clock stability by as much as a factor of 10, compared to a typical CC that makes use of fluorescence detection.

Absorption is another way of detecting the signal in a CC.  However, many practical issues must be taken into account if absorption is to be used.  First, the fluctuation in the clock frequency is affected by additional noise contributed by the laser used in absorption.  Let us assume that the observation time window is $\tau$, and the number of photons in the probe beam, before absorption, is $N_P$, and the probe is in a Coherent state.  We also assume that the number of atoms passing through the detection process in this time window is $N_A$, and the linewidth of the resonance is $\Gamma$.  If the detection process produces an absorption by a fraction of $\alpha$ (i.e., $\alpha=1$ represents perfect absorption of the laser beam), and the detector has a quantum efficiency of $\eta$, then the resulting fluctuation in the clock frequency can be expressed as:
\begin{equation}
\delta\omega_{ABS}=\Gamma\left( \frac{1}{\sqrt{\eta\alpha N_A}}+\frac{1}{\sqrt{\eta\alpha N_P}} \right)
\label{absorption}
\end{equation}

Here, the first term inside the parenthesis represents the quantum projection noise of the atoms, and the second term represents the shot noise of the photons (which can be thought of as the quantum projection noise of photons). The validity of this expression can be easily established by considering various limits. Consider first the ideal case where $\xi \equiv \eta\alpha=1$. For $N_P \gg N_A$, the additional noise from the laser can be neglected, and we get the fundamental noise limit due to the quantum projection noise of atoms.  On the other hand, if $N_A \gg N_P$, the quantum projection noise from the atoms can be neglected, and the process is limited by the shot-noise of the laser.  In general, the parameter $\xi$ represents the overall quantum efficiency of the detection process.  The corresponding expression for detection via fluorescence is $\delta\omega_{FLU}=\Gamma(\eta\rho N_A)^{-1/2}$, where again $\eta$ is the quantum efficiency of the detector, and $\rho$ is the fraction of fluorescence falling on the detector.

The contribution from the second term in Eq. \eqref{absorption} shows that the intensity of the laser beam used in absorption must be made strong enough in order to make the effect of this term negligible compared to the first term.  However, since the absorption process is nonlinear and saturates for a strong laser beam, increasing the laser intensity often decreases the effective value of $\alpha$. For example, consider an ensemble of $2 \cdot 10^6$ atoms with a linear resonant optical density of ~300, which can be realized (as we have shown above) for an ensemble confined to a cigar shaped ensemble with a diameter $\sim 50$ $\mu$m.  For a weak probe, the value of $\alpha$ is unity.  However, as the probe power is increased, the value of $\alpha$ decreases dramatically.  This can be seen by considering a situation where the value of $N_P$ is $10^9$, for example.  Since the atomic transition used for absorption is not closed (i.e., not cyclic), the ensemble can only absorb a number of photons that is of the order of $2\cdot 10^6$.  Thus, the maximum value of $\alpha$ would be only about 0.002.  Furthermore, if the area of the laser beam ($A_L$) is much larger than the area of the atomic ensemble ($A_A$), then the value of $\alpha$ can never exceed the value of $A_A/A_L$.  We are not aware of any publication reporting a cold atom clock that makes use of absorption for detecting the atoms, possibly because of these constraints and considerations.  Nonetheless, as a matter of principle, an absorption process can certainly be used to reduce the quantum frequency fluctuation below what is observed in fluorescence detection systems, under proper choice of parameters..

\section{Physical Interpretation of Linewidth Reduction and Its Relevance to the Transit Time Limit}
\label{sec:interpret}
As we have shown, the fact that the linewidth in a COSAC is narrower by a factor of $\sqrt{N}$ can be proven mathematically. However, it is instructive to discuss the physical mechanism that leads to this narrowing. Furthermore, it is also important to address the issue of why the violation of the conventional notion of the transit time limit does not contradict the fundamental laws of quantum mechanics. 

\subsection{Physical interpretation of line narrowing}

We consider a simple picture of an oscillator and a probe in order to understand the physical explanation as to why the linewidth of a COSAC narrows by $\sqrt{N}$. A clock is essentially an oscillator oscillating at some frequency $\omega$. In order to ascertain that the oscillator has not drifted, the oscillator frequency is mapped into light and interacts with a two level atom, with the ground state $|1\rangle$ and the excited state $|2\rangle$,  and a transition frequency $\omega_0$. If $\omega$ does not match $\omega_0$, an error signal proportional to $\delta=\omega-\omega_0$ is produced to correct for the difference. Now consider for a moment that we can create a two state superposition of $N$ atoms such that they are all either in the ground state or the excited state. In other words, $|\psi\rangle=C_0|E_0\rangle+C_N|E_N\rangle$ where $|E_0\rangle=|111...11\rangle$ and $|E_N\rangle=|222...22\rangle$. The energy difference between these two states is $N\omega_0$. The oscillator frequency is still $\omega$, but when a light field with $N$ photons is compared with such a two level system, the difference in energy is $N\delta=N\omega-N\omega_0$. If it were possible to produce an error signal that is proportional to this energy difference without degrading the effective signal to noise ratio (or, more accurately, the ratio of noise to the SVS, as discussed in Section V-A), the resulting clock would be $N$-fold more accurate. This is functionally equivalent to the clock transition frequency being enhanced by a factor of N.

\begin{figure}
	\includegraphics[width=.3\textwidth]{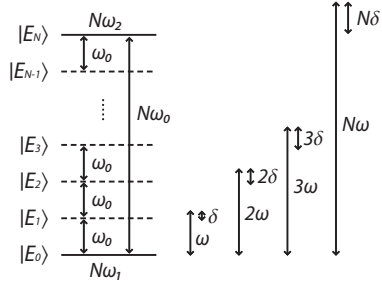}
	\caption{Collective state energy levels, separated by $\omega_0$, are excited by a field of frequency $\omega$. All the states from $|E_0\rangle$ to $|E_N\rangle$ are excited, and participate in producing an effective clock transition frequency proportional to $\sqrt{N}$.}
	\label{fig:phys_n}
\end{figure}

However, this clean two level superposition of collective states is virtually impossible to achieve with a collection of $N$ non-interacting atoms and a single field since there is no electric dipole moment to excite the $|E_N\rangle$ state directly from the $|E_0\rangle$ state. What occurs instead is that all the states between these get excited as well, as illustrated in Fig. \ref{fig:phys_n}. If we consider only the excitations from state $|E_0\rangle$, there are $N$ possible transitions that can occur, so that the error signal includes the set of all the possible detunings, $\delta, 2\delta, 3\delta, ... N\delta$. In other words, there are effectively $N$ different sensors running at the same time. All the other states also act as sensors as they interact with the others. It turns out, as we have proven mathematically in Section \ref{sec:collective}, that the error signal becomes proportional to $\sqrt{N}\delta$, corresponding to an effective detuning of $\sqrt{N}\delta$. This is functionally equivalent to the clock transition frequency being enhanced by a factor of $\sqrt{N}$. 

\begin{figure}
	\includegraphics[width=.45\textwidth]{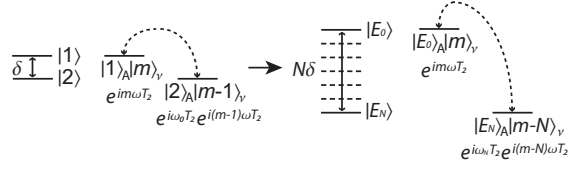}
	\caption{Ramsey fringe experiment of a two level atom, in the Jaynes Cummings model, involves states $|1\rangle_A |m\rangle_\nu$ and $|2\rangle_A |m-1\rangle_\nu$ where the state with subscript $A$ represents the atomic state, and subscript $\nu$ represents the Ramsey field. The phase difference of the two levels at the end of the experiment is $e^{i\delta T_2}$, and the signal produced would oscillate at frequency $\delta$; If a two level system existed in which the ground state were the collective state $|E_0\rangle_A|m\rangle_\nu$ and the excited state were the collective state $|E_N\rangle_A|m-N\rangle_\nu$, the phase accumulation between the two states at the end of the Ramsey fringe experiment would be $e^{iN\delta T_2}$, and the oscillation frequency would be $N\delta$. }
	\label{fig:rams}
\end{figure}

In the Ramsey fringe experiment, the error signal that is generated occurs as a result of the phase difference between the interacting states. A detailed picture can be viewed in Fig. \ref{fig:rams}.  Consider first a single two level atom, initially in state $|1\rangle_A$, going through the Ramsey fields. In the Jaynes-Cummings model, when a field with $m$ photons interacts with an atom, the $\pi/2$-pulse will produce the quantum state $|\psi\rangle=|1\rangle_A|m\rangle_\nu-i|2\rangle_A|m-1\rangle_\nu$. The energy of state $|2\rangle_A|m-1\rangle_\nu$ is lower than that of state $|1\rangle_A|m\rangle_\nu$ by $\hbar\delta$. In the second zone, these two composite states evolve freely for a time $T_2$ and accumulate different phases. State $|1\rangle_A$, with energy $0$ remains the same, whereas $|2\rangle_A$ with energy $\omega_0$ evolves as $e^{i\omega_0 T_2}$. The field with $m$ photons evolve as $e^{im\omega T_2}$ whereas the field with $m-1$ photons evolve as $e^{i(m-1)\omega T_2}$. Thus, the quantum state of the total system at the end of the dark zone is
\begin{equation}
|\psi\rangle=e^{im\omega T_2}|1\rangle_A|m\rangle_\nu -i e^{i\omega_0 T_2}e^{i(m-1)\omega T_2}|2\rangle_A|m-1\rangle_\nu
\end{equation}
The net accumulated phase difference in the two states is $e^{i\delta T_2}$. The third zone where another $\pi/2$-pulse occurs produces interference between the two states, so that when interrogation occurs, the signal produced is in the form of Ramsey fringes that oscillate at frequency $\delta$. Therefore, the energy difference between the two composite states determines the oscillation frequency of the Ramsey fringes. Alternatively, if one were to plot the signal as a function of the dark zone time, $T_2$, the width of the fringe is given by the inverse of this energy difference. If the same calculation is carried out now for a two state system where the ground state is $|E_0\rangle_A|m\rangle_\nu$ and the excited state is $|E_N\rangle_A|m-N\rangle_\nu$, where $|E_0\rangle_A$ and $|E_N\rangle_A$ are the collective states of $N$ atoms, then the energy difference is $N\delta$ and the width of the fringe as a function of $T_2$ would be $1/(N\delta)$ and the width of the Ramsey fringe as a function of $\delta$ will be $(T_2^{-1}/N)$.

As mentioned earlier, such a two level system of collective states for a large value of $N$ is virtually impossible to realize for non-interacting atoms. Instead, for $N$ atoms, the first Ramsey zone produces a superposition of all the states from $|E_0\rangle_A$ to $|E_N\rangle_A$. In the second zone, each of the collective states  $|E_k\rangle_A$ accumulates a phase factor of $e^{i (\delta T_2)k}$ with respect to the state $|E_0\rangle_A$. When the atoms pass through the third zone, each of these collective states interferes with one another and contributes to the total population of $|E_N\rangle_A$. It is the collection of these interferences among all the collective states that produces the narrowed linewidth. 

\begin{figure}
	\includegraphics[width=.48\textwidth]{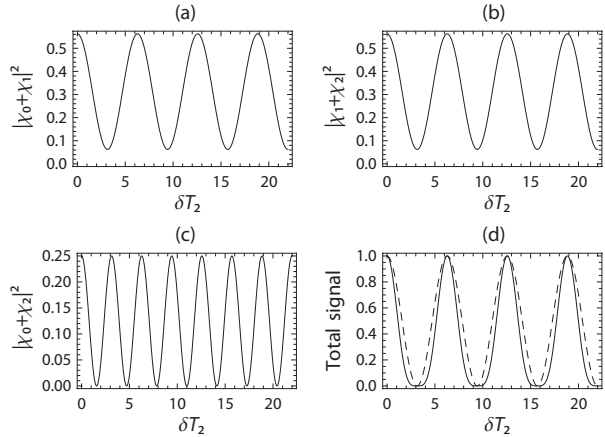}
	\caption{In a two atom ensemble, each of the three collective states interfere with one another to produce different Ramsey fringes (a)-(c). The overall envelope is not drawn. The sum of these interferences gives the narrowing of the fringe linewidth as seen in (d). In (d), the dotted curve represents the signal from a single atom and the solid curve the signal from two atoms for comparison.}
	\label{fig:phys}
\end{figure}

We have verified this interpretation explicitly for two atoms. The collective states in this case are (where the subscript $A$ has been dropped) $|E_0\rangle$, $|E_1\rangle$, and $|E_2\rangle$. After they accumulate different phases in the second zone, each of them contributes  to the final state $|E_2\rangle$ by amount $\chi_0=1/4$, $\chi_1=e^{i\delta T}/2$, and $\chi_2=e^{2i\delta T}/4$ respectively. The total signal is $S_{col}=|\langle E_2|E_2\rangle|^2=\cos^4{(\delta T_2/2)}$. This comes about because $S_{col}=|\chi_0+\chi_1+\chi_2|^2=|\chi_0+\chi_1|^2+|\chi_1+\chi_2|^2+|\chi_0+\chi_2|^2-(\chi_0^2+\chi_1^2+\chi_2^2)$. In other words, it is as though $|E_0\rangle$ and $|E_1\rangle$ interfered together to produce Ramsey fringes at frequency $\delta$, $|E_1\rangle$ and $|E_2\rangle$ interfered together to produce Ramsey fringes at frequency $\delta$, and $|E_0\rangle$ and $|E_2\rangle$ interfered together to produce Ramsey fringes at frequency $2\delta$; the signal observed is the addition of all these Ramsey fringes minus an overall factor (see Fig. \ref{fig:phys}), which is due to the fact that the actual process is a simultaneous interference between the three states.

\subsection{Violation of the conventional notion of the transit time limit}
The narrowing of the COSAC fringe as given by $\Gamma(N)=\Gamma(1)/\sqrt{N}=\pi/(T_2\sqrt{N})$ violates the conventional transit time limit, which constrains the fringe width to be at least $\sim 1/T_2$. This is a manifestation of the uncertainty relation $\Delta f \cdot \Delta t \geq 1$, which apparently follows from the Heisenberg uncertainty principle of $\Delta E \cdot \Delta t \geq \hbar$. However, when we properly define $\Delta f$ as the uncertainty in the fringe width -- in the case of the Ramsey technique considered here -- and $\Delta t$ as the total observation time, we can derive the uncertainty relations more systematically and show that despite the fact that the conventional transit time limit is violated, the Heisenberg uncertainty principle is not violated. 

First, consider a single atom that undergoes the Ramsey fringe experiment. The uncertainty in the fringe width is $\Delta f=(1/T_2)$, where $T_2$ is the separation period between the two $\pi/2$ pulses. When the experiment is repeated $m$ times, it is as though the separation period expands $m$-fold, so that the effective observation time is in fact $\Delta t=mT_2$, and the uncertainty in the fringe width is $\Delta f=(1/T_2)/\sqrt{m}$ in the standard quantum limit (SQL) and $\Delta f=(1/T_2)/m$ in the Heisenberg limit (HL). Hence, the product $\Delta f \cdot \Delta t$ yields $\sqrt{m}$ in the SQL and $1$ in the HL. Note that as $m\rightarrow 1$, the SQL approaches the HL, which is the more fundamental limit. 

Next, consider $N$ atoms in the same Ramsey fringe experiment during a single trial. Since each atom, in its individual state, is considered separately from the rest, having $N$ atoms is equivalent to running $N$ trials simultaneously. The effective observation time in this case is $\Delta t=NT_2$, and the uncertainties in the fringe width are $\Delta f=(1/T_2)/\sqrt{N}$ in the SQL and $\Delta f=(1/T_2)/N$ in the HL. Moreover, if the experiment is repeated $m$ times, the effective observation time increases to $\Delta t=mNT_2$, and the uncertainties in the fringe width are $\Delta f=(1/T_2)/\sqrt{mN}$ in the SQL and $\Delta f=(1/T_2)/(mN)$ in the HL. Thus, we find that the uncertainty relations for $N$ atoms and $m$ trials are $\Delta f\cdot \Delta t=\sqrt{mN}$ in the SQL and $\Delta f\cdot \Delta t=1$ in the HL. 

Consider next the COSAC case, containing $N$ atoms, and repeated $m$ times. As we have shown in Section \ref{sec:performance}, the frequency fluctuation in the COSAC is $\Delta f = 1/(T_2\sqrt{mN})$ for ideal detection efficiency. It may not be obvious what the effective observation time is for this case. However, given the fact that, under ideal detection efficiency, the COSAC is equivalent to the case of $N$ atoms repeated $m$ times, we are led to conclude that the effective observation time is $\Delta t = T_2mN$. As such, we get $\Delta f\cdot \Delta t=\sqrt{mN}$, which is the SQL in this case. In the HL, we could get $\Delta f\cdot \Delta t=1$. Thus, we see that when the frequency uncertainty and the observation times are interpreted properly, the COSAC signal does not violate the fundamental quantum limit.

\section{Conclusion}
\label{sec:conclusion}

We have described an atomic clock with a significant reduction in the Ramsey fringe linewidth, by a factor of $\sqrt{N}$, by measuring the amplitude of a collective state with a heterodyne detection scheme. We have shown that the reduction occurs due to multipath interference among the collective states, and does not violate the fundamental quantum limit. The performance of the COSAC has been compared to that of the CC by analyzing quantum and classical fluctuations in frequency. When the effects of detector efficiency and collection efficiency are considered, it can be seen that the COSAC may perform 10 times better than a typical CC employing fluorescence detection.

\begin{acknowledgments}
This work has been supported by the NSF grants number DGE-0801685 and DMR-1121262 , and AFOSR grant number FA9550-09-1-0652.
\end{acknowledgments}

\bibliography{cosac}

\end{document}